\documentclass[pre,showpacs,twocolumn,amsmath,amssymb,floatfix,superscriptaddress]{revtex4-1}

\usepackage[utf8]{inputenc}
\usepackage{graphicx}
\usepackage{color}
\usepackage{bbold}

\definecolor{darkgreen}{rgb}{0,0.7,0}

\graphicspath{{Figures_loc/}}

\newcommand{\dt}{\Delta\tau}

\newcommand{\dT}{\Delta\tau}
\newcommand{\refs}[1]{Sec.~\ref{sec:#1}}
\newcommand{\refss}[1]{Sec.~\ref{subsec:#1}}

\newcommand{\reff}[1]{Fig.\ \ref{fig:#1}}
\newcommand{\reffa}[1]{Fig.\ \ref{fig:#1}(a)}
\newcommand{\reffb}[1]{Fig.\ \ref{fig:#1}(b)}
\newcommand{\reffc}[1]{Fig.\ \ref{fig:#1}(c)}
\newcommand{\reffd}[1]{Fig.\ \ref{fig:#1}(d)}
\newcommand{\reffl}[1]{Figure\ \ref{fig:#1}}
\newcommand{\refq}[1]{(\ref{eq:#1})}
\newcommand{\myparagraph}[1]{{\it #1} -- }

\newcommand{\wn}{i\omega_n}

\makeatletter
  \def\l@subsubsection#1#2{}%
\makeatother

\begin{document}
	\title{Quasi-continuous-time impurity solver for the dynamical mean-field theory\\ 
	with linear scaling in the inverse temperature}

\author{D.~Rost}
\affiliation{Institute of Physics, Johannes Gutenberg University, Mainz, Germany}
\affiliation{Graduate School Materials Science in Mainz, Johannes Gutenberg University, Mainz, Germany}
\author{F.~Assaad}
\affiliation{Institute of Theoretical Physics and Astrophysics, University of W\"urzburg, W\"urzburg, Germany}
\author{N.~Bl\"umer}
\affiliation{Institute of Physics, Johannes Gutenberg University, Mainz, Germany}
\date{\today}
  \begin{abstract}     
  We present an algorithm for solving the self-consistency equations of the dynamical mean-field theory (DMFT) with high precision and efficiency at low temperatures. In each DMFT iteration, the impurity problem is mapped to an auxiliary Hamiltonian, for which the Green function is computed by combining determinantal quantum Monte Carlo (BSS-QMC) calculations with a multigrid extrapolation procedure. The method is 
numerically exact, i.e., yields results which are
free of significant Trotter errors, but retains the BSS advantage, compared to direct QMC impurity solvers, of linear (instead of cubic) scaling with the inverse temperature. 
The new algorithm is applied to the half-filled Hubbard model close to the Mott transition; detailed comparisons with exact diagonalization, Hirsch-Fye QMC, and continuous-time QMC are provided.
  \end{abstract}
  \pacs{02.70.Ss, 71.10.Fd, 71.27.+a}
  \maketitle


\section{Introduction}\label{sec:Intro}

\vspace{-1ex}
The dynamical mean-field theory (DMFT)\cite{Metzner1989,Krauth1996,Kotliar2006,Vollhardt2012} and its cluster extensions \cite{Kotliar2001,Maier2005a} are powerful approaches for the numerical treatment of correlated electron systems, both in the model context and for materials science, e.g., embedded in the LDA+DMFT \cite{Held2006} or GW+DMFT \cite{Biermann2003,Held2011,Taranto2012b,Tomczak2012} frameworks which extend density functional theory to strongly correlated materials \cite{Kotliar2004,Kotliar2006}. 
Recently, many DMFT studies have also appeared in the context of ultracold fermions on optical lattices \cite{Joerdens2010,Gorelik2010,Gorelik2012}.
The DMFT reduces electronic lattice models to impurity problems, which have to be solved self-consistently \cite{Jarrell1992,Georges1992,Janis1992}.
A challenging part of this iterative procedure is the computation of the interacting Green function for a given impurity configuration (defined by the fixed local interactions and the self-consistent Weiss field). Thus, the availability of efficient and reliable {\it impurity solvers} determines the complexity of models and the parameter space that can be accessed using the DMFT.

Quantum Monte Carlo (QMC) impurity solvers allow for numerically exact solutions of the DMFT self-consistency equations at finite temperatures. In the case of the Hirsch-Fye auxiliary field (HF-QMC) method \cite{Hirsch1986,Jarrell1992,Blumer2011a}, all raw estimates contain systematic errors due to the inherent Trotter decomposition and associated imaginary-time discretization \cite{Hirsch1986,Fye1987}; unbiased results can only be obtained after an extrapolation of the discretization interval $\dt\to 0$ \cite{Blumer2005a,*Blumer2005b,Blumer2007}. Diagrammatic QMC impurity solvers \cite{Rubtsov2005,Werner2006a,Werner2006b,Gull2011} sample partition function and Green functions in continuous (imaginary) time (CT), i.e., avoid systematic biases. However, in all of these direct QMC approaches, the computational effort scales cubically \cite{Gull2007} with the inverse temperature $\beta=1/k_{\text{B}}T$, which limits their access to low-temperature phases.

Exact diagonalization (ED) based impurity solvers \cite{Caffarel1994} require a discrete representation of the impurity action in terms of an auxiliary Hamiltonian, which is then solved either by full diagonalization (for evaluations at arbitrary temperature) or using a Lanczos procedure \cite{Koch2011} (e.g., at $T=0$). As the numerical effort scales exponentially with the number $N_b$ of auxiliary ``bath'' sites, $N_b$ has to be kept quite small, which introduces, again, a bias 
and is a particularly severe limitation for multi-orbital or cluster DMFT studies at finite temperatures.

Recently, Khatami {\it et al.} proposed another Hamiltonian-based scheme \cite{Khatami2010a}, in which the Green function and other relevant properties of the auxiliary problem are computed using the determinantal BSS-QMC method developed by Blankenbecler, Scalapino, and Sugar \cite{Blankenbecler1981}. 
The advantage of this scheme, compared to ED, is the possibility of using more bath sites (due to cubic instead of exponential scaling with $N_b$); the advantage over the direct QMC impurity solvers is the linear, instead of cubic, scaling in $\beta$ \cite{Khatami2010b}. The authors established the feasibility of the method 
and proved that the associated sign problem (arising at general band filling in cluster extensions of DMFT, in frustrated lattices, and for generic multiband models) converges to that of HF-QMC for sufficiently fine bath discretization \cite{Khatami2010a}. However, as all BSS-QMC applications to date, the Green functions and all observable estimates resulting from their implementation suffer from systematic Trotter errors.

In this work, we construct a similar algorithm where the Trotter bias inherent in the BSS Green functions is eliminated using a multigrid procedure before feeding them back in the self-consistency cycle. As a DMFT building block, the resulting method is an exact quasi-CT QMC impurity solver with linear scaling in the inverse temperature. Its scaling advantage over direct QMC impurity solvers should allow access to lower temperatures, in particular in multi-orbital and cluster DMFT studies.

The paper is organized as follows: In \refs{methods}, we briefly review the DMFT equations and the BSS-QMC algorithm and fully specify our multigrid DMFT-BSS approach. As a test case, the new method is applied (in single-site DMFT) to the half-filled Hubbard model in the vicinity of the Mott transition in \refs{results}. Here, we first focus on the Green function at moderately low temperature $T=t/25$ and then discuss important observables, namely the double occupancy and quasiparticle weight, also at lower temperatures. The accuracy of our approach is established by comparisons with the results of the (multigrid) HF-QMC, ED and CT-QMC solvers, as well as with the previous (finite $\dt$) DMFT-BSS implementation. We show that our elimination of the Trotter error improves the results dramatically. We also discuss the impact of the bath discretization and establish convergence to the thermodynamic limit. A summary and outlook conclude the paper in \refs{conclusion}.



\section{Theory and Algorithms}\label{sec:methods}

In this section, we lay out the proposed algorithm for solving the DMFT self-consistency equations without significant Trotter errors and with a computational effort that grows only linearly with the inverse temperature.
We start out by reviewing the general DMFT framework and established methods (ED, HF-QMC, CT-QMC) for its solution in \refss{DMFT} in sufficient detail to expose the similarities and differences with respect to the new method. Here we also discuss some algorithmic choices, in particular regarding the Hamiltonian representation in our ED implementation, that are essential ingredients also for the BSS-QMC based approaches. 
We then turn to the BSS-QMC method and its applicability in the DMFT context in \refss{BSS}, and specify, finally, our new numerically exact implementation in \refss{multigrid}.
For simplicity, and in line with the numerical results to be presented in \refs{results}, we write down the formalism for the single-band Hubbard model and the original, single-site variant of DMFT. Extensions to cluster DMFT (and to multiband models) should be straightforward, but require some generalizations (e.g. for the treatment of offdiagonal Green functions) and will be pursued in a subsequent publication.

\subsection{DMFT and established impurity solvers}\label{subsec:DMFT}

\myparagraph{The Hubbard model on a lattice or graph}
We consider the single-band Hubbard model
\begin{equation}\label{eq:Hubbard}
  H = H_0 + H_{\text{int}} = \sum_{ij,\sigma}t_{ij}\,c_{i\sigma}^\dagger c_{j\sigma}^{\phantom{\dagger}} + U \sum_i n_{i\uparrow}n_{i\downarrow}\,,
\end{equation}
where $c_{i\sigma}^\dagger$ ($c_{i\sigma}^{\phantom{\dagger}}$) creates (annihilates) an electron with spin $\sigma \in \{ \uparrow, \downarrow  \} $ on lattice site $i$; $n_{i\sigma}=c_{i\sigma}^\dagger c_{i\sigma}^{\phantom{\dagger}}$ is the corresponding density, $t_{ij}=t_{ji}$ the hopping amplitude between sites $i$ and $j$ (or the local potential for $i=j$); $U$ quantifies the on-site interaction. Usually, the hopping is defined to be translationally invariant, e.g., $t_{ij}=-t$ for nearest-neighbor bonds on an infinite mathematical lattice [as illustrated in \reffa{mapping} for a square lattice] 
or on a finite cluster with periodic boundary conditions. However, neither the DMFT nor direct QMC approaches to the Hubbard model depend crucially on such assumptions, as will be discussed in \refss{BSS}.

\myparagraph{General DMFT self-consistency procedure}
If all lattice sites are equivalent and for spatially homogeneous phases, the DMFT maps the original lattice problem \refq{Hubbard}, illustrated in \reffa{mapping}, onto a single-impurity Anderson model [\reffb{mapping}], which has to be solved self-consistently.
The impurity problem is defined by its action
\begin{align}
  \label{eq:action}
  \mathcal{A}\left[ \psi, \psi^*, \mathcal{G} \right] &= \int\limits_0^\beta\int\limits_0^\beta d\tau\, d\tau^\prime\, \sum_\sigma \psi^*_\sigma(\tau)\,\mathcal{G}^{-1}_\sigma\, \psi^{\phantom{*}}_\sigma(\tau^\prime) \\ \notag
  &- U \int \limits_0^\beta d\tau\, \psi^*_\uparrow(\tau)\, \psi^{\phantom{*}}_\uparrow(\tau)\, \psi^*_\downarrow(\tau)\, \psi^{\phantom{*}}_\downarrow(\tau) \text{ ,}
\end{align}
here in imaginary time $\tau\in [0,\beta]$ and in terms of Grassmann fields $\psi$, $\psi^*$. ${\cal G}$ is the ``bath'' Green function, i.e., the noninteracting Green function of the impurity, which is related to the full impurity Green function $G$, 
\begin{equation}\label{eq:green_def}
	G_{\sigma}(\tau) = 
	- \langle \mathcal{T}_\tau\, \psi_{\sigma}^{\phantom *}(\tau)\, \psi_{\sigma}^*(0) \rangle_{\cal A}
\end{equation}
(with the time ordering operator $\mathcal{T}_\tau$), 
and the self-energy $\Sigma$ by the (impurity) Dyson equation
\begin{equation}
\label{eq:dyson1}
  G_\sigma^{-1}(\wn) = \mathcal{G}_\sigma^{-1}(\wn) - \Sigma_\sigma (\wn)\text{ ,}
\end{equation}
here written in terms of fermionic Matsubara frequencies $\omega_n =(2n+1)\pi T$ at finite temperature $T$; here and in the following, we set $\hbar=k_{\text{B}}=1$.
\begin{figure} 
	\unitlength0.1\columnwidth
	\begin{picture}(10,3.6)
	  \put(0.0,0.0){\includegraphics[width=\columnwidth]{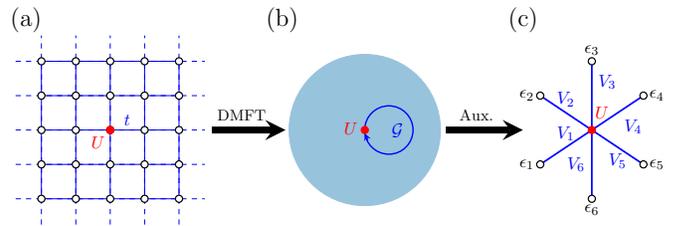}}
	\end{picture}
\caption{(Color online) Mapping of the original lattice problem (a), with local interaction $U$ and hopping $t$, on a single impurity (b), embedded in an effective bath ${\cal G}$.
(c) Discretization of the bath in terms of an auxiliary Hamiltonian (treatable with ED or BSS-QMC), here with star topology. \label{fig:mapping}}
\end{figure}
\begin{figure*}
        \begin{center}
         \includegraphics[width=\textwidth]{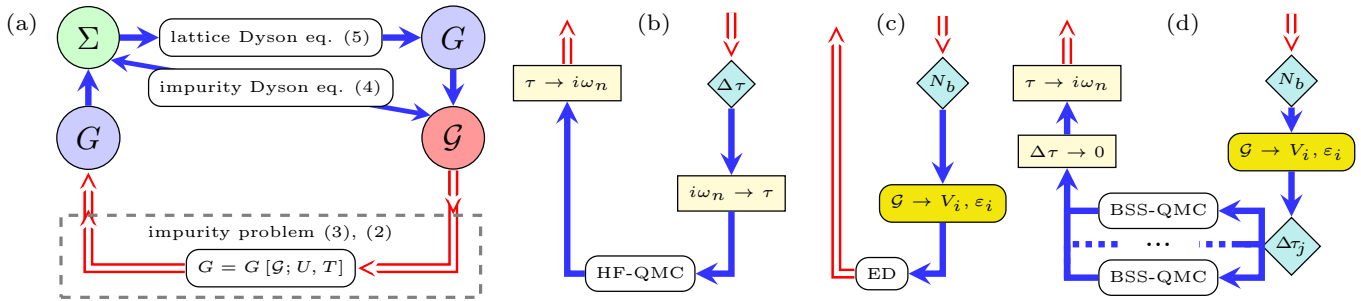}
        \end{center}
        \caption{(Color online) (a) Scheme of the general DMFT self-consistency cycle, including the ``impurity problem'' (dashed box). Established impurity solvers include (b) the Hirsch-Fye (HF-QMC) algorithm and (c) exact diagonalization (ED): cf.\ the main text. 
(d) The proposed algorithm approximates the bath Green function ${\cal G}$ in terms of the parameters $V_i$, $\epsilon_i$ of an auxiliary Hamiltonian \refq{H_anderson} with $N_b$ ``bath'' sites [like ED (c)]. Corresponding Green functions are computed using BSS-QMC for a grid $\dt_{\text{min}}\le \dt_j\le \dt_{\text{max}}$ of Trotter discretizations.
The subsequent extrapolation of $\dt\to 0$  yields the Green function, free of significant Trotter errors and continuous in $\tau$, which is easily Fourier transformed and fed back into the self-consistency cycle.}
        \label{fig:DMFT_cycle}
\end{figure*}   

The central DMFT assumption is that of a local self-energy on the lattice \cite{Metzner1989}: $\Sigma_{ij\sigma}(\wn)=\delta_{ij}\Sigma_{ii\sigma}(\wn)$, which is identified with the impurity self-energy. Similarly, the impurity Green function is identified with the local component of the lattice Green function:
\begin{align}
	G_\sigma(\wn) &= G_{ii\sigma}(\wn)  = \left\{ \mathbb{t} +  \left[ \wn + \mu -\Sigma_\sigma(\wn)\right] \mathbb{1} \right\}^{-1}_{ii}\nonumber\\
	&= \int_{-\infty}^\infty d \varepsilon \,\frac{\rho (\varepsilon)}
	{i \omega_n +  \mu - \Sigma_{\sigma}(\wn)- \varepsilon}\,, \label{eq:Dyson}
\end{align}
where the last expression is valid in the homogeneous case, $\rho (\varepsilon)$ denotes the noninteracting density of states, and $\mathbb{t}$ is the matrix with elements $t_{ij}$.

The general DMFT iteration scheme is illustrated in \reffa{DMFT_cycle}: starting, e.g., with an initial guess $\Sigma=\Sigma_0$ of the self-energy, the Green function $G$ is computed using the lattice Dyson equation \refq{Dyson}. In a second step, $\Sigma$ and $G$ yield the bath Green function ${\cal G}$ via the impurity Dyson equation \refq{dyson1}, which defines, in combination with the local interactions, the impurity problem [illustrated in \reffb{mapping}], the solution of which is the nontrivial part of the algorithm. A second application of the impurity Dyson equation \refq{dyson1}, to the resulting $G$ and to ${\cal G}$, yields a new estimate of the self-energy $\Sigma$, which closes the self-consistency cycle. In the following, we discuss the primary options for addressing the impurity problem.

\myparagraph{Direct impurity solvers}\label{subsec:direct}
One class of methods directly evaluates the path integral representation of the Green function [Eqs.\ \refq{green_def} and \refq{action}] for a continuous bath ${\cal G}$, 
which corresponds to a DMFT solution of the original lattice problem in the thermodynamic limit after self-consistency. We will refer to such methods as ``direct impurity solvers.''

For a long time, the Hirsch-Fye QMC (HF-QMC) algorithm has been the method of choice for nonperturbative DMFT calculations \cite{Jarrell1992}. HF-QMC is based on a discretization of the imaginary time $\tau\in [0,\beta]$ into $\Lambda$ ``time slices'' of width $\dt=\beta/\Lambda$, a Trotter decomposition of the interaction and kinetic terms in Eq.\ \refq{action}, and a Hubbard-Stratonovich transformation, which replaces the electron-electron interaction by an auxiliary binary field on each time slice; the resulting problem is then solved employing Wick's theorem and Monte Carlo importance sampling over the field configurations. As configurations can be updated in the case of a single spin flip (i.e., an auxiliary-field change on a single time slice) with a matrix-vector operation of cost ${\cal O}(\Lambda^2)$ and $\Lambda$ local updates are needed for a global configuration update, the numerical cost of the HF-QMC algorithm scales as $\Lambda^3$. All HF-QMC results have statistical errors (which decay as $N^{-1/2}$ for $N$ ``sweeps,'' each consisting of $\Lambda$ attempted single-spin updates) and systematic errors resulting from the Trotter decomposition. As $\dt$ has to be kept constant for roughly constant systematic error upon variation of $T$, the numerical effort of HF-QMC scales as the cube of the inverse temperature, $\beta^3$. This is also true for the numerically exact (unbiased) ``multigrid'' HF-QMC method \cite{Blumer2008,*Gorelik2009}.

The integration of the (conventional) HF-QMC method into the DMFT self-consistency cycle is illustrated in \reffb{DMFT_cycle} [as a specification of the lower dashed box in \reffa{DMFT_cycle}]: a fixed choice of $\dt$ (diamond-shaped selection box) defines the grid $\tau_l=l\dt$ with $0\le l\le \Lambda$ for a Fourier transform (square box) of the Matsubara bath Green function ${\cal G}(\wn)$ (with $|\omega_n|\le \omega_\text{max}$ for some cutoff frequency $\omega_\text{max}$) to the imaginary-time equivalent $\{{\cal G}(\tau_l)\}_{l=0}^\Lambda$. After application of the HF-QMC algorithm (rounded box), the result $\{ G(\tau_l)\}_{l=0}^\Lambda$ is transformed back to Matsubara frequencies (square box); this step requires special care in order to get around the Nyquist theorem, e.g. using analytic weak-coupling results \cite{KnechtMaster,BlumerThesis,Blumer2007}.

More recently, conceptionally different QMC approaches have been formulated, which are based on diagrammatic expansions of the action \refq{action} in continuous imaginary time, either in the interaction $U$ (CT-INT \cite{Rubtsov2004}) or in the bath hybridization (CT-HYB \cite{Werner2006a,Werner2006b}),  and on a stochastic sampling of Feynman diagrams;  CT-AUX \cite{Gull2008} is related to the HF-QMC method \cite{Gull2011}.  All of these continuous time (CT-QMC) algorithms require Fourier transforms, before and (with the exception of CT-INT) after the QMC part; the numerical cost is associated primarily with matrix updates, similar to those arising in HF-QMC, with a total scaling of the computational effort, again, as $\beta^3$. 

Thus, all direct QMC-based impurity solvers are very costly at low $T$, which limits their access to low-temperature phases of particular physical interest.

\myparagraph{Auxiliary Hamiltonian and exact diagonalization}
Another class of numerical approaches, such as the ``exact diagonalization'' methods, cannot directly be applied to the action-based formulation of the impurity problem, but requires a Hamiltonian representation \cite{Koch2008}. One possibility is the ``star topology'' illustrated in \reffc{mapping}, where a central ``impurity'' site (with the same interactions as the impurity problem, here $U$) is connected by hopping matrix elements $V_{i\sigma}$ to a number $N_b$ of noninteracting ``bath sites,'' each characterized by a local potential $\epsilon_{i\sigma}$. In general, this representation has to be spin-dependent, leading to the Anderson Hamiltonian
\begin{align}
  H_\text{And} = \ &\epsilon_0 \sum_\sigma n_\sigma  + U n_{\uparrow}n_{\downarrow} \notag \\ 
                  & + \sum_\sigma  \sum_{i=1}^{N_b} \left [\epsilon_{i\sigma} n_{i\sigma} + V_{i\sigma} \left( a^\dagger_{i\sigma} c_\sigma^{\phantom{\dagger}}  +  \text{h.c.} \right) \right]\text{ ,}\label{eq:H_anderson}
\end{align}  
where $c_\sigma^\dagger$ ($c_\sigma^{\phantom{\dagger}}$) creates (annihilates) an electron with spin $\sigma \in \{ \uparrow, \downarrow  \} $ on the impurity site and $a_{i\sigma}^\dagger$ ($a_{i\sigma}^{\phantom{\dagger}}$) creates (annihilates) an electron with spin $\sigma$ on bath site $i$; $n_{\sigma}=c_\sigma^\dagger c_\sigma^{\phantom{\dagger}}$, $n_{i\sigma}=a_{i\sigma}^\dagger a_{i\sigma}^{\phantom{\dagger}}$ are the corresponding number operators. 
In this work, we consider only nonmagnetic phases, which implies spin symmetric bath parameters $V_{i\downarrow}=V_{i\uparrow}$, $\epsilon_{i\downarrow}=\epsilon_{i\uparrow}$.

For a fixed choice of $N_b$,
the bath parameters $\epsilon_{i\sigma}, V_{i\sigma}$ are determined such that the noninteracting impurity Green function $\mathcal{G}_{\text{And},\sigma}$ associated with $H_\text{And}$, with
\begin{equation}
\label{eq:G_anderson}
  \mathcal{G}_{\text{And},\sigma}^{-1}   \left( \omega  \right) = \omega +\mu_\sigma - \sum_{i=1}^{N_b} \frac{V_{i\sigma}^2}{\omega - \epsilon_{i\sigma}} \text{,}
\end{equation}
 is ``close'' to the target bath Green function ${\cal G}$ according to some metric (see below). 
Note that the resulting spectrum $-\frac{1}{\pi}\text{Im}\,\mathcal{G}_{\text{And},\sigma}(\omega+i0^+)$ is necessarily discrete (for finite $N_b$), in contrast to piece-wise smooth spectrum of the true bath Green function; in this sense, the mapping to a Hamiltonian implies a ``bath discretization'' in frequency space; this step clearly introduces a bias which has to be controlled particularly carefully within iterative procedures such as the DMFT.

The integration of this type of approach in the DMFT cycle is illustrated for the case of ED in \reffc{DMFT_cycle}: for a fixed choice of $N_b$ (diamond shaped box), the parameters $V_i, \epsilon_{i}$ (here and in the following we suppress spin indices) are adjusted (rounded box) as to minimize the bath misfit
\begin{equation}
  \label{eq:chi2}
  \chi^2\left[ \{ V_{i}, \varepsilon_{i} \}\right] = \sum_{n=0}^{n_c} \text{w}_n \left| \mathcal{G_\text{And}}(\wn;\{ V_{i}, \varepsilon_{i} \}) - \mathcal{G}(\wn) \right|^2 \text{,}
\end{equation}
with a cutoff Matsubara frequency $i\omega_{n_c}$ and the weighting factor w$_n$, which can be used to optimize the bath parametrization \cite{Senechal2010} and which we set to w$_n=1$%
\ \footnote{In our experience, this choice for w$_n$ is not crucial: other choices lead to very similar results.}.
As this fit is performed directly on the Matsubara axis, no Fourier transform is needed for ${\cal G}$. Using ED (rounded box), the Green function $G$ can be evaluated on the Matsubara axis \cite{Caffarel1994}; therefore, the DMFT cycle is closed without any Fourier transform. 

The minimization of $\chi^2$ [as defined in Eq.\ \refq{chi2}] is performed in our ED and BSS-DMFT calculations using the Newton method, based on analytic expressions for the derivative $\nabla \chi^2$ with respect to the bath parameters. 
Due to the multidimensional character of the problem, this deterministic method is often trapped in local minima; thus, a naive implementation of Newton based methods will, in general, not find globally optimal parameters, which can induce unphysical fixed points in the DMFT iteration procedure.
Therefore, we use not only the solution $\{V_i, \epsilon_i\}$ of the previous iteration as initialization, but perform a large number (up to 1000) of independent Newton searches, starting also from random initial parameters. Of the resulting locally optimal solutions, we choose the one with minimum $\chi^2$ as the final result of the minimization procedure; typically, about $1\%$ of the individual searches come close to this (estimated) global optimum.

An advantage of ED, compared to QMC algorithms, is that Green functions and spectra can be computed directly on the real axis, without analytic continuation; however, numerical broadening of the resulting discrete peaks is required. This discretization problem is particularly severe as the numerical effort of the matrix diagonalization scales exponentially with the total number of sites (here $N_b+1$), which limits the applicability of ED for cluster extensions of DMFT or multiband models. 


\subsection{Principles of the BSS-QMC algorithm and application as a DMFT impurity solver}\label{subsec:BSS}
In Eq.\ \refq{H_anderson}, we have used the conventional notation for the auxiliary Hamiltonian that emphasizes its interpretation as an impurity model, e.g., with different creation operators for electrons on the central ``impurity'' site ($c_\sigma^\dagger$) and on the bath sites ($a_{i\sigma}^\dagger$), respectively. However, with the changes $c_\sigma \to c_{0\sigma}$, $n_\sigma\to n_{0\sigma}$, $a^\dagger_{i\sigma}\to c^\dagger_{i\sigma}$, and $V_{i\sigma}\to t_{0i}^\sigma$, it essentially reproduces the Hubbard model \refq{Hubbard} on a graph, just with nonuniform interaction ($U$ acting only on site 0) and, possibly, spin-dependent hopping amplitudes and local energies. 

As a consequence, the model \refq{H_anderson} is not only treatable with the universal ED approach, but also with more specific methods developed for Hubbard-type models. As pointed out recently by Khatami \textit{et al.} \cite{Khatami2010a}, this includes the determinantal quantum Monte Carlo approach by Blankenbecler,  Scalapino,  and Sugar \cite{Blankenbecler1981,Assaad08_rev}
 (BSS-QMC), which, thereby, becomes applicable as a DMFT impurity solver. In the following, we will first sketch the established BSS-QMC approach (for an extended discussion, including issues of parallelization, see Ref.~\onlinecite{Bai2010}) and then discuss its application in the DMFT context.

Similarly to the HF-QMC method (cf. \refss{direct}), the BSS-QMC approach is based on a 
Trotter-Suzuki decomposition, here of the partition function
\begin{align}
  \label{eq:trotter_decomp}
     Z =&\, \text{Tr} \left( e^{-\beta (H_K + H_V)} \right) \\\label{eq:trotter_decomp2}
       \approx& \, Z_{\dt} =  \text{Tr} \left( \prod_{l=0}^\Lambda e^{-\dT H_K}e^{-\dT H_V} \right)  \text{,}
\end{align}
where $H_V$ ($H_K$) corresponds to the interaction (kinetic and local potential) contribution to the Hubbard type models \refq{Hubbard} or \refq{H_anderson} and $\dt=\beta/\Lambda$. 
Again, a discrete Hubbard-Stratonovich transformation replaces the interaction term by a binary auxiliary field $\{ h \}$ with $h_i(l)=\pm 1$ at each site $i$ and time slice $l$.  The trace in Eq.\ \refq{trotter_decomp2} then simplifies to
\begin{align}\label{eq:trotter_decomp3}
  Z_{\dt} &= \sum_{\{ h \}} \text{det} \left[ M_\uparrow^{\{ h \}} \right] \text{det}\left[ M_\downarrow^{\{ h \}} \right]  \quad \mbox{with} \\ \notag
    M_\sigma^{\{ h \}} &= \mathbb{1} + B_{\Lambda,\sigma}\left[\{ h_i (\Lambda)\}_{i=1}^N\right]\, \ldots\, B_{1,\sigma}\left[\{ h_i (1)\}_{i=1}^N\right] \text{,}
\end{align}
where $B$ is defined in terms of the hopping matrix $K$: 
\begin{equation}
  B_{l,\sigma}\left[ \{ h_i (l)\}_{i=1}^N \right] = e^{\sigma \lambda \text{diag}\left[ h_1(l), \ldots, h_N (l) \right]}\, e^{- \dT K} \text{.}
\end{equation}
The interaction strength is encoded in the parameter $\lambda=\text{cosh}^{-1}(e^{U \dT /2})$. 
The computation of thermal averages of physical observables $O$ takes the form:
\begin{align}
  \left\langle O \right\rangle &=  \sum_{\{ h \}} \left[O^{\{ h \}} \ \mathcal{P}_{\dT}^{\{ h \}} \right] \\ \notag  
                              \mathcal{P}_{\dT}^{\{ h \}} &= \frac{1}{Z_{\dT}} \text{det}\left[ M_\uparrow^{\{ h \}} \right] \text{det}\left[ M_\downarrow^{\{ h \}} \right] \text{ .}
\end{align}
At particle-hole symmetry, the weights $\mathcal{P}_{\dT}^{\{ h \}}$ are always positive; i.e., the sums can be evaluated at arbitrary precision, without any sign problem.
As in HF-QMC, the problem is solved by Monte Carlo importance sampling of the auxiliary field $\{h\}$ and evaluation of the Green function at time slice $l$, with
\begin{equation}
  \label{eq:G_B}
  G_{l,\sigma}^{\{ h \}} = \left[ \mathbb{1} + B_{l-1,\sigma}\, \ldots \, B_{1,\sigma} B_{\Lambda,\sigma}\, \ldots \, B_{l,\sigma}\right]^{-1} \text{.}
\end{equation}
As a spin flip in the auxiliary field $h_i(l)$ at time slice $l$ and site $i$ only affects $B_{l,\sigma}$ at this site, the ratio of the weights, needed for the decision whether a proposed spin flip is accepted, involves only local quantities; a full recomputation of the determinants of $N \times N$ matrices appearing in Eq.\ \refq{trotter_decomp3} is not needed. The  computational effort is further reduced by calculating the Green function at time slice $l+1$ from the quantities at time slice $l$, using so-called ``wrapping'':
\begin{equation}
  \label{eq:G_update}
  G_{l+1,\sigma} = B_{l,\sigma}^{-1} G_{l,\sigma} B_{l,\sigma} \text{ .}
\end{equation}
In order to avoid the accumulation of numerical errors in the matrix multiplications, it is necessary to recalculate the full Green function at regular intervals. This is particularly important at low temperatures.

All this considered, the numerical cost scales cubically with the number of sites and linearly with the number of time slices; at constant $\dt$, this translates to a total effort $\mathcal{O}(N^3 \beta)$, where $N=N_b+1$. 
Note that a need for finer bath discretizations at lower temperatures could potentially spoil the scaling advantage of the method over direct impurity solvers; we will show in \refss{results_lowT} that this is not the case for our test applications.

The application in the DMFT context \cite{Khatami2010a} starts with the computation of the Hamiltonian parameters (for some choice of $N_b$), exactly like in the ED approach. As in the HF-QMC approach, one then chooses some discretization $\dt$, computes $\{G(l\dt)\}_{l=0}^\Lambda$ for the impurity site, and applies a (nontrivial) Fourier transform back to Matsubara frequencies. The result is an impurity solver with superior scaling (linear in $\beta$) compared to the direct impurity solvers (cubic in $\beta$), however with a bias due to the Trotter discretization $\dt$ (in addition to a possible bias due to the bath discretization with  $N_b$ sites) which, as we will show in \refs{results}, can be quite significant.


\subsection{Specification of multigrid BSS-QMC algorithm}\label{subsec:multigrid}
The central feature of our new algorithm is the elimination of this systematic Trotter error, while retaining the advantage of linear-in-$\beta$ scaling inherent in the BSS-QMC method. 
In the following, we will specify the method and illustrate it using an example (\reff{Gtau_m_i_1}), that will be discussed in detail in \refs{results}.

In contrast to the previous DMFT-BSS implementation with a unique discretization $\dT$ in all BSS computations throughout the DMFT self-consistency cycle, the (impurity) Green function of the Hamiltonian $H_\text{And}$ at hand is computed in $M\gtrsim 20$ parallel BSS runs (indexed by $1\le i \le M$), each employing a homogeneous imaginary-time grid with a specific discretization $(\dT)_i$, chosen from a set $\{(\dT)_i| (\dT)_\text{min} \leq (\dT)_i \leq (\dT)_\text{max}\}$ with typically 6 - 9 different elements. Green functions resulting from BSS-QMC runs with the same discretization $(\dT)_i=(\dT)_j$ are averaged over, thereby reducing the dependencies on initialization conditions and further enhancing the parallelism.

  This leads to a set of Green functions defined, in general, on incommensurate imaginary-time grids (symbols in \reff{Gtau_m_i_1}). In order to apply a local $\dT \rightarrow 0$ extrapolation, all $G_{(\dT)_i}$ have to be transformed to a common grid. This is possible since the true $G(\tau)$ is a smooth function; however, a direct spline interpolation of the raw QMC results, neglecting higher derivatives, would not be accurate \cite{Joo2001}. Instead, we consider differences between the raw data $\{G_{(\dT)_i}(l\dt)\}_{l=0}^{\Lambda_i}$ and a reference Green function, obtained via Eq.\ \refq{Dyson} from a model self-energy $\Sigma_\sigma^{\text{ref}}(\wn)$ \cite{KnechtMaster,BlumerThesis}, written here for the single-band case (for multi-band generalizations, see Ref. \cite{Knecht2005}):
\begin{align}
  \Sigma^\text{ref}_\sigma(\wn) &= U \left( \left\langle n_{-\sigma} \right\rangle - \frac{1}{2}\right)\notag +\frac{1}{2}U^2\left\langle n_{-\sigma} \right\rangle \left(1-\left\langle n_{-\sigma} \right\rangle\right)\\ 
   &\times \left( \frac{1}{\wn+\omega_0} + \frac{1}{\wn-\omega_0}\right)\text{ ,}\label{eq:model_SE}
\end{align}
which recovers the exact high-frequency asymptotics of $\Sigma(\wn)$ and $G(\wn)$ for any choice of the free parameter $\omega_0$ and, therefore, approximates the second and higher-order derivatives of $G(\tau)$ at $\tau\to 0$ (and $\tau\to \beta$) well. This match can be further improved by adjusting $\omega_0$. Consequently, the differences $\{G_{(\dT)_i}(l\dt) - G^{\text{ref}}(l\dt)\}_{l=0}^{\Lambda_i}$ have smaller absolute values and much smaller higher derivatives than the original data; in particular, their curvature vanishes asymptotically at the boundaries 
\footnote{A reasonably good approximation of the curvature of $G(\tau)$ at the boundaries by the reference Green function  is the only crucial requirement for this procedure; it can also be achieved using other model self-energies or, alternatively, via maximum entropy analytic continuation of the ``best'' measured $G_{(\dT)_i}$.}.
Thus, they are well represented by natural cubic splines.

Usually, the parameters of the piecewise polynomials constituting such a spline $f_{\text{spline}}(x)$ are determined from discrete data $\{f_{\text{meas}}(x_i)\}_{i=0}^N$ such that the discrete data are reproduced exactly: $f_{\text{spline}}(x_i)=f_{\text{meas}}(x_i)$ for all $0\le i\le N$. However, in the QMC context, all measurements have statistical errors, i.e., the discrete data are better represented as $\{f_{\text{meas}}(x_i) \pm \Delta f_{\text{meas}}(x_i)\}_{i=0}^N$ with standard deviations $\Delta f_{\text{meas}}$, which are also estimated within the QMC procedure. It is clear that the usual interpolating splines, which do not take the uncertainties of the discrete data into account, contain more features than warranted by the data (in particular at the Nyquist frequency); in the context of Green functions this includes the possibility of acausal behavior. We use, instead, smoothing spline fits  \cite{Cox1978,Enting2006} which reproduce the discrete data only within error bars, which are typically ${\cal O} (10^{-3})$, (and minimize the curvatures under this constraint); these fits can be computed in a very similar procedure and at the same cost as interpolating splines. 

After combining these approximating ``difference'' splines with exact expressions for $G^{\text{ref}}(\tau)$ resulting from Eqs.\ \refq{model_SE} and \refq{Dyson}, we obtain smooth approximations of the Green functions, as seen in \reffa{Gtau_m_i_1}; 
\begin{figure} [t] 
	\unitlength0.1\columnwidth
	\begin{picture}(10,13.8)
	  \put(0.0,0.0){\includegraphics[width=\columnwidth]{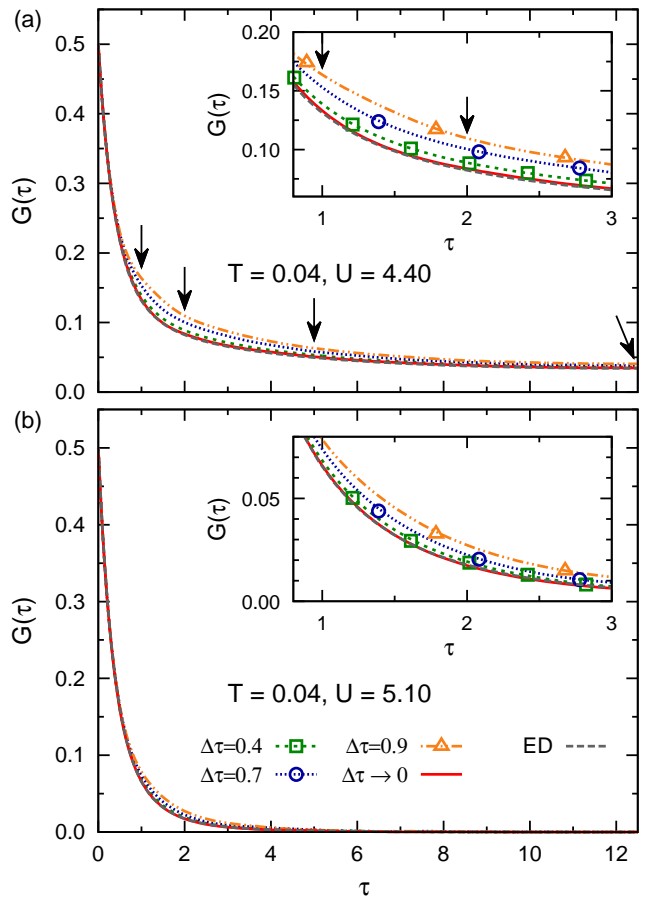}}
	\end{picture}
	\caption{(Color online) BSS-QMC impurity Green functions at $T=0.04$ (symbols) using a bath representation with $N_b=4$ sites  (with parameters of converged DMFT-ED solution, long-dashed lines) and results of multigrid extrapolation  to $\dT = 0$ (solid lines). Upper panel: metallic phase ($U=4.4$). Lower panel: insulating phase ($U=5.1$). Arrows denote $\tau$ values for which the extrapolation is shown in \reff{Gtau_extra}.}
	\label{fig:Gtau_m_i_1}
\end{figure}
the inset also demonstrates slight deviations of the continuous spline fits from the discrete data (within error bars), e.g., for the discretization $\dt=0.7$ (dotted line) at $\tau\approx 1.4$ (circle), while most other data points are reproduced within the line widths.

These smooth approximations can be evaluated on an arbitrarily fine common grid (e.g. with $\dt_\text{fine}=0.005$) and extrapolated to $\dt\to 0$. This is illustrated in \reff{Gtau_extra} for the representative values of $\tau$ denoted by arrows in \reffa{Gtau_m_i_1}. Even though most of the raw BSS-QMC data do not include estimates of the Green functions at these precise values of $\tau$, the transformed data (symbols in \reff{Gtau_extra}) depend very regularly on $\dt$, falling on nearly straight lines as a function of $(\dt)^2$. Therefore, they can accurately and reliably be extrapolated to $\dt\to 0$ (lines in \reff{Gtau_extra} and symbols at $\dt=0$); an application of this procedure at all $\tau$ (on the fine grid) leads to quasi-continuous Green functions without significant Trotter errors, shown as solid lines in \reff{Gtau_m_i_1}. These results can be Fourier transformed to Matsubara frequencies in a straightforward manner [cf. \reffd{mapping}]. A similar approach has also been useful for computing unbiased spectra from BSS-QMC \cite{Rost2012}.
\begin{figure} [t]  
	\includegraphics[width=\columnwidth]{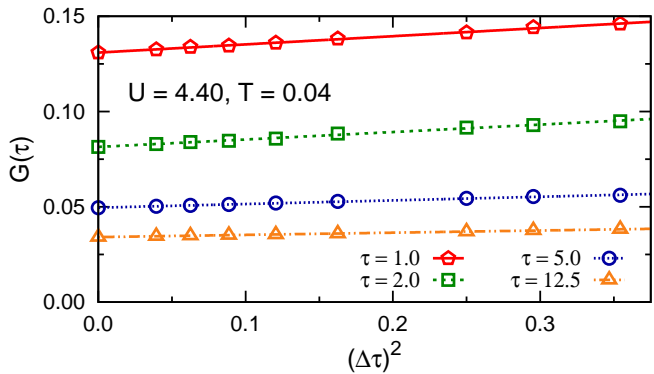}
	\caption{(Color online) BSS-QMC estimates of imaginary-time Green functions $G_{\dt}(\tau)$ at $T=0.04$, $U=4.4$ after interpolation (corresponding to the colored broken lines in \reff{Gtau_m_i_1})
	 for selected values of $\tau$ (symbols) and extrapolation to $\dt=0$ using least-squares fits (lines).}
	\label{fig:Gtau_extra}
\end{figure}

At first sight, the computational advantage of the multigrid procedure is less obvious in the BSS-QMC context than for HF-QMC \cite{Blumer2008,*Gorelik2009,Blumer2012}, since the numerical effort for direct computations at small $\dt$ grows only linearly, not cubically, with $(\dt)^{-1}$ in the BSS case (while the systematic errors decay generically as $(\dt)^2$ for a given impurity problem). However, even for a fixed Hamiltonian, so much accuracy can be gained by extrapolation that it more than offsets the cost of the additional grid points. This is true, in particular, since stable results are best obtained by averaging over independent BSS-QMC runs; performing these on variable grids then allows for extrapolation without additional cost. Furthermore, the individual runs thermalize faster in the multigrid variant, 
due to the smaller number of time slices (and proportionally shorter run time per sweep), which enhances the parallelism.
Most importantly, as we will see below, the DMFT self-consistency can magnify any bias of the employed impurity solvers in complicated ways (in the vicinity of phase transitions), so that controlled results are really dependent on unbiased methods, such as our multigrid approach.

\section{Results}
\label{sec:results}
In this section, we compare results of the new numerically exact ``multigrid'' BSS-QMC method with raw BSS-QMC results (at finite Trotter discretization), with reference ED results (which are exact at the level of the auxiliary Hamiltonian), and with the predictions of established impurity solvers (multigrid HF-QMC \cite{Blumer2008,*Gorelik2009} and CT-HYB \cite{Werner2006a,Werner2006b,Alps_2_0}). These comparisons are performed in three stages: In \refss{results_fixed_H}, we keep the bath ${\cal G}$ and its approximation by an auxiliary Hamiltonian fixed and discuss the impact of the Trotter error and its elimination without the complications of the DMFT self-consistency. In \refss{results_Trotter}, we compare full DMFT solutions obtained using the various algorithms at moderate temperature ($T=0.04$), focusing on the impact of Trotter errors on the resulting estimates of double occupancy and quasiparticle weight. Finally, we present results also at lower temperatures $T\ge 0.01$ (with DMFT self-consistency), where the impact of the bath discretization becomes particularly relevant, in \refss{results_lowT}.

Following the established practice for the evaluation of DMFT impurity solvers \cite{Gull2007,Blumer2007}, all of these comparisons are performed for the half-filled Hubbard model with semi-elliptic ``Bethe'' density of states \cite{Kollar2005} (full band width $W=4$) within the paramagnetic phase. Specifically, we choose temperatures $T\le 0.04$, which are below the critical temperature $T^*\approx 0.055$ \cite{BlumerThesis,Blumer2012} of the first-order metal-insulator transition, and interactions close to or within the coexistence region of metallic and insulating solutions, which arises from the mean-field character of the DMFT.

\subsection{Green function extrapolation at fixed bath Hamiltonian parameters}
\label{subsec:results_fixed_H}

In general, a bias present in an impurity solver has a two-fold impact: On the one hand, it affects estimates of Green functions and all other properties for a given impurity problem, defined by its bath Green function ${\cal G}$. On the other hand, it shifts the fixed point of the DMFT self-consistency cycle, i.e., it also modifies the converged bath Green function, which, in turn, also affects the measured Green functions and all other properties. In this subsection, we study the first effect in isolation by fixing the bath Green function to the converged solution of the ED procedure for $N_b=4$ bath sites (with Hamilton parameters $\{\epsilon_i, V_i\}_{i=1}^4$). As the same auxiliary Hamiltonian is used also in the BSS-QMC algorithm, the ED estimates of the Green function are exact for the purpose of the current comparison; the impact of the bath discretization (which corresponds to a bias on the DMFT level) will be discussed in \refss{results_lowT}.

Local imaginary-time Green functions $G(\tau)$ are shown in \reffa{Gtau_m_i_1} for the metallic phase, at $U=4.4$, and in \reffb{Gtau_m_i_1} for the insulating phase, at $U=5.1$. Here and in the following, we restrict the imaginary-time range to $0\le \tau\le \beta/2$; data for $\tau>\beta/2$ follow from the particle-hole symmetry $G(\beta-\tau)=G(\tau)$. Symbols (in the magnified insets) represent raw BSS-QMC results (with discretizations $\dt=0.4$, $\dt=0.7$, and $\dt=0.9$); colored long-dashed, dotted, and dash-dotted lines denote interpolations obtained using the methods described in \refss{multigrid}. Due to the large discretization, these data deviate significantly from the ED reference results (gray long-dashed lines), in particular at moderately low imaginary times $\tau\approx 2$. In contrast, multigrid BSS-QMC Green functions (solid lines) are indistinguishable from the ED data at $U=5.1$ and very close to them at $U=4.4$, with deviations of the order of statistical errors. Thus, our method yields, indeed, quasi-continuous Green functions without significant Trotter errors in both test cases, although the discretizations of the underlying raw BSS-QMC computations (with $0.3\le \dt \le 1.0$) would be considered much too coarse in conventional applications. 

\begin{figure} [t] 
    \includegraphics[width=\columnwidth]{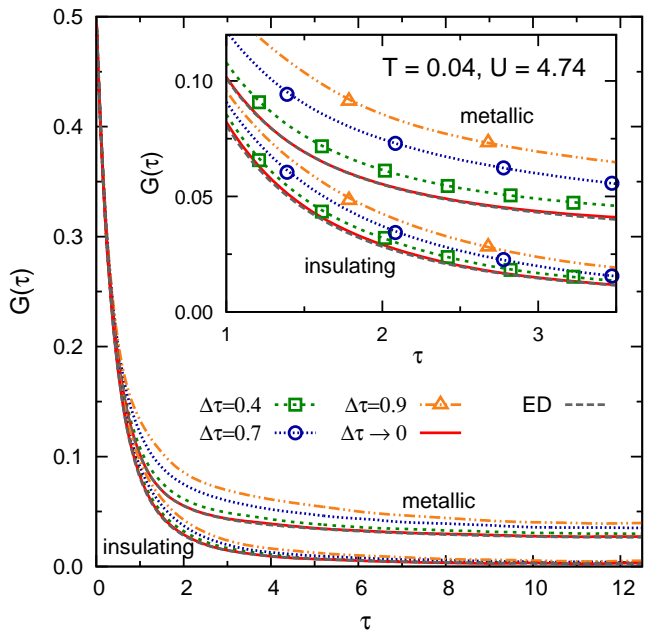}
	\caption{(Color online) BSS-QMC impurity Green functions at $T=0.04$ and $U=4.74$ (symbols and colored broken lines) using a bath representation with $N_b=4$ sites  (with parameters fixed by converged DMFT-ED solution, long-dashed lines) and extrapolation  to $\dT = 0$ (solid lines). Upper (lower) set of curves: metallic (insulating) bath.}
	\label{fig:Gtau_m_i_2}
\end{figure}
A very similar picture emerges in an analogous comparison for the two coexisting solutions at $U=4.74$, shown in  \reff{Gtau_m_i_2}. Again, the raw BSS-QMC results (symbols and colored broken lines) show a strong systematic bias, towards more metallic Green functions and of different magnitude in the different phases, while the extrapolated Green functions agree nearly perfectly with the ED references. In fact, some of the BSS Green functions calculated for an insulating bath (lower set of symbols and broken lines) show such large discretization errors at small $\tau\lesssim 2$, that they approach the exact Green function of the metallic DMFT solution (upper solid and long-dashed lines). One may suspect from this observation that these biased ``insulating'' solutions will not be associated with stable DMFT fixed points if they are fed back in the self-consistency cycle; such shifts of stability regions induced by the Trotter bias at $\dt>0$ will, indeed, be seen in \refss{results_Trotter}.

\begin{figure} [t] 
  \includegraphics[width=\columnwidth]{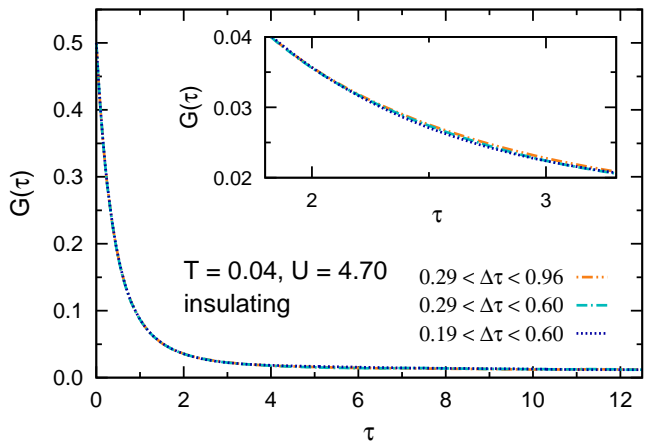}
	\caption{(Color online) Green functions in the insulating phase at $T=0.04$, $U=4.7$, extrapolated from BSS-QMC results using different imaginary-time grids [53] 
	 (at fixed bath represen\-tation). The excellent agreement shows that the multigrid procedure is stable with respect to its technical parameters. 	}
	\label{fig:Gtau_U47B25Nb4dts}
\end{figure}
It is clear that the proposed multigrid extrapolation technique can only be useful as a practical method if it is insensitive to the particular set $\{(\dT)_i \}$ of discretizations in the underlying BSS-QMC runs, i.e., if no sensible choice leads to a significant bias. This is demonstrated in \reff{Gtau_U47B25Nb4dts} for the insulating phase at $U=4.7$: the Green functions for the same auxiliary problem obtained from multigrid extrapolations with three different $\dT$ grids 
\footnote{The results shown in \reff{Gtau_U47B25Nb4dts} correspond to the following grids of the discretization $\dt$: [0.29, 0.40, 0.50, 0.60, 0.69, 0.78, 0.89, 0.96] (yellow line), [0.29 0.34 0.40 0.44 0.50 0.54 0.60] (light blue line), and [0.19 0.25 0.29 0.34 0.40 0.44 0.50 0.54 0.60] (dark blue line).}
agree perfectly within the precision of the method.
The latter is primarily determined by the statistical errors, i.e., by the number of sweeps and, possibly, by the numerical precision in the matrix operations. Only if raw BSS-QMC data of much higher precision was available (with many millions of sweeps per run), additional accuracy could be gained by choosing smaller discretizations (e.g., $0.1\le \dt\le 0.3$). As a rule of thumb, the multigrid procedure can be based on discretizations $\dt$ that are 3 to 10 times as large as the discretization that one would choose in a conventional BSS-QMC procedure.

\subsection{Comparisons of impurity solvers at full DMFT self-consistency: impact of Trotter errors}
\label{subsec:results_Trotter}

So far, we have compared different algorithms just at the impurity level, i.e., for a fixed bath Green function (determined from a self-consistent DMFT-ED calculation). In contrast, we will now discuss results of completely independent DMFT solutions, each of which corresponds to full self-consistency for a given impurity solver (cf. \reff{DMFT_cycle}). For all Hamiltonian based methods (ED, BSS-QMC, and multigrid BSS-QMC), the number of bath sites is restricted to $N_b=4$ (as above); the impact of this parameter will be studied in \refss{results_lowT}.

Specifically, we discuss static observables that are particularly useful for 
discriminating between metallic and (possibly coexisting) insulating DMFT solutions, namely
the double occupancy
\begin{equation}
  \label{eq:DO}
   D = \langle n_{\uparrow} n_{\downarrow}  \rangle\,,
\end{equation}
which is proportional to the interaction energy $E_{\text{int}}=UD$,
and the quasiparticle weight
\begin{equation}
  \label{eq:Z}
   Z = \left[ 1  -  \frac{\partial\, \text{Re}\,\Sigma(\omega)}{\partial \omega}\Big|_{\omega=0}\right]^{-1}   \!
\approx \left[ 1  +  \frac{\text{Im}\,\Sigma(i\omega_1)}{\pi T}\right]^{-1}\,.
\end{equation}

Open symbols in \reff{D_Z_B25} denote estimates resulting from self-consistent DMFT solutions using the conventional BSS-QMC impurity solver at finite discretization $0.3\le \dt \le 0.5$, i.e., using the scheme established in Ref. \cite{Khatami2010a}. 
\begin{figure} [t] 
  \unitlength0.1\columnwidth
  \begin{picture}(10,10)
    \put(0,0){\includegraphics[width=\columnwidth]{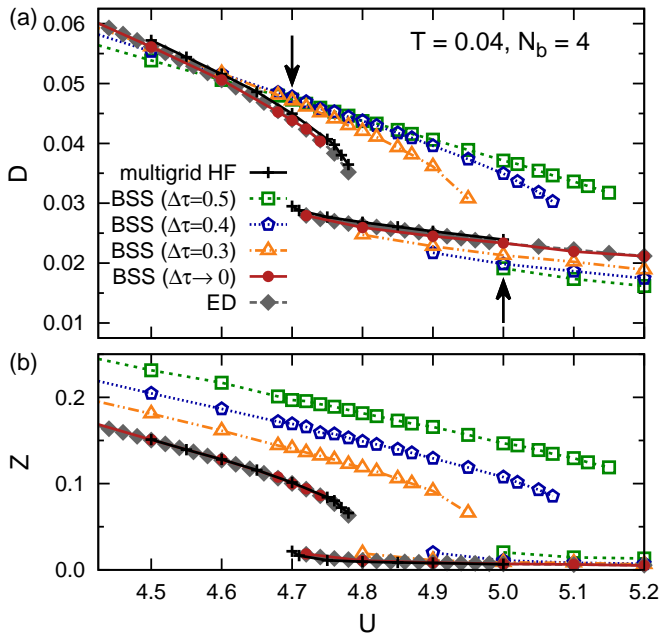}}
  \end{picture}
	\caption{(Color online) Estimates of double occupancy $D(U)$ and quasiparticle weight $Z(U)$ obtained in independent selfconsistent DMFT calculations using various impurity solvers: multigrid HF-QMC (crosses), conventional BSS-QMC (open symbols), multigrid BSS-QMC (circles), and ED (diamonds). In each panel, the upper (lower) sets of curves correspond to metallic (insulating) solutions. Lines are guides to the eye only. Arrows in (a) indicate parameters for which the discretization dependence is studied in \reff{D_extrapol}.}
	\label{fig:D_Z_B25}
\end{figure}
The estimated values of $Z$, shown in \reffb{D_Z_B25}, have a nearly uniform offset in the metallic phase at $U\lesssim 4.8$ relative to each other and relative to the reference ED solution (gray diamonds). The Trotter bias inherent in the conventional BSS-QMC procedure also leads to a significant overestimation of the range of stability of the metallic solution: The metallic BSS-QMC solutions extend to much larger interactions (e.g., to $U\approx 5.1$ at $\dt=0.4$) than the ED reference solution.

This is also seen in corresponding estimates of the double occupancy [\reffa{D_Z_B25}]; however, for these observables the Trotter bias is highly nonuniform (in the metallic solution): at $U=4.7$ (arrow), the conventional BSS-QMC estimates are nearly on top of each other; relative deviations are only clearly seen at stronger interactions $U\gtrsim 4.9$ and (to a lesser degree) at weaker interactions $U\lesssim 4.5$. At the same time, nearly all of these data deviate significantly (and without obvious systematics) from the reference ED result (diamonds), so that an {\it a posteriori} elimination of the Trotter bias seems impossible.

In contrast, the new multigrid BSS-QMC procedure, as discussed in \refss{multigrid} and illustrated in \reffd{DMFT_cycle}, leads to estimates of both $D$ and $Z$ (filled circles) which perfectly recover the ED solutions, even though they are based on BSS-QMC runs with $\dt\ge 0.3$. 

This is also true for the insulating solutions (lower sets of curves in \reff{D_Z_B25}), the stability range of which is also shifted towards stronger interactions in the case of conventional BSS-QMC calculations (open symbols); here, the Trotter bias appears roughly uniform for $D$ and very nonuniform for $Z$. Again, the multigrid BSS-QMC results agree perfectly with the ED reference data.

For comparison, crosses and black solid lines in \reff{D_Z_B25} denote estimates of an unbiased direct impurity solver, namely the multigrid HF-QMC method \cite{Blumer2008,*Gorelik2009}; these show good overall agreement with both the ED and the multigrid BSS-QMC data. A slight  negative deviation in the estimates of $D$ of the latter, Hamiltonian based, methods can be traced back to the relatively poor bath discretization with $N_b=4$ auxiliary sites (cf.\ \refss{results_lowT}). 

Since the double occupancy $D$ is best computed directly on the impurity level (in QMC based approaches), its physical value has to be extrapolated from raw estimates $D_{\dt}$, with discretizations corresponding to the different grid points used within the multigrid procedure  (in contrast to the quasiparticle weight $Z$, which follows from the self-energy $\Sigma$, which, in turn, is determined from unbiased Green functions). 
\begin{figure} [t] 
  \includegraphics[width=\columnwidth]{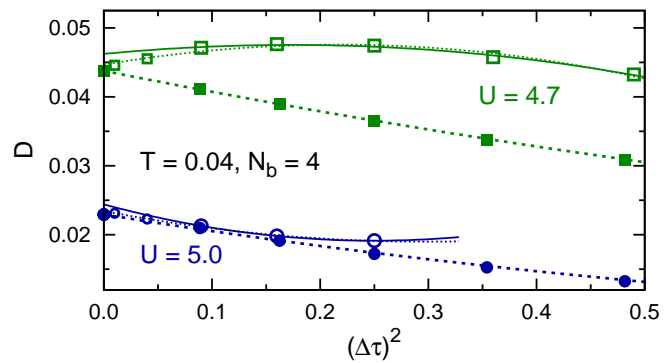}
	
\caption{(Color online) Discretization dependence of the double occupancy $D$ as estimated from BSS-QMC, using either the multigrid scheme (filled symbols) or self-consistent BSS-QMC solutions at finite $\dt$ (open symbols), within the metallic phase at $U=4.7$ (upper data set) or the insulating phase at $U=5.0$ (lower data set). Dashed (solid) lines denote least squares fits to the multigrid (conventional) BSS-QMC data at $\dt\ge 0.3$; dotted lines denote fits that include also data at $\dt=0.1$ and $\dt=0.2$ (small open symbols).}	\label{fig:D_extrapol}
\end{figure}

As seen in \reff{D_extrapol}, the Trotter bias inherent in these raw estimates (filled symbols) is perfectly regular\
\footnote{Specifically, we have used fit laws of the form $\ln[D(\dt)] = D(\dt=0) + A (\dt)^2$.}
even at large $\dt$, so that reliable extrapolations $\dt\to 0$ (thick dashed lines) are possible both in the metallic phase, at $U=4.7$ (upper set of curves), and in the insulating phase, at $U=5.0$ (lower set of curves).

In contrast, estimates of $D$ resulting from conventional BSS-QMC calculations in the same range of discretizations $\dt\ge 0.3$ (large open symbols in \reff{D_extrapol}) show such irregular dependencies on $\dt$ that quadratic least-square fits (solid lines) lead to extrapolations $\dt\to 0$ with significant offsets. Roughly accurate results (dotted lines) can only be obtained when including raw data at much smaller discretizations (small open symbols). This shows, again, that only an elimination of all Trotter errors {\it within} the self-consistency cycle, as introduced by our multigrid approach, can efficiently generate high-precision results.

\subsection{Comparisons of impurity solvers at full DMFT self-consistency: impact of bath discretization}
\label{subsec:results_lowT}
So far, we have restricted the bath representation in all Hamiltonian based impurity solvers (ED and both variants of BSS-QMC) to only $N_b=4$ bath sites and focused on the impact of the Trotter errors and their elimination. From the mutual agreement with multigrid HF-QMC, an impurity solver which treats the bath directly on the action level, we can conclude that this coarse bath discretization allows for reasonably accurate estimates of $D$ and, in particular, $Z$ at the moderately low temperature $T=0.04$. However, the ED and multigrid BSS-QMC estimates of $D$ were found in \reff{D_Z_B25} to lie a bit below the multigrid HF-QMC data; this deviation must be an artifact of the bath discretization if the multigrid HF-QMC reference data are correct. Moreover, we must suspect that the bath discretization bias gets worse (at constant $N_b$) at lower temperatures.

\begin{figure} [t] 
  \unitlength0.1\columnwidth
  \begin{picture}(10,10)
    \put(0,0){\includegraphics[width=\columnwidth]{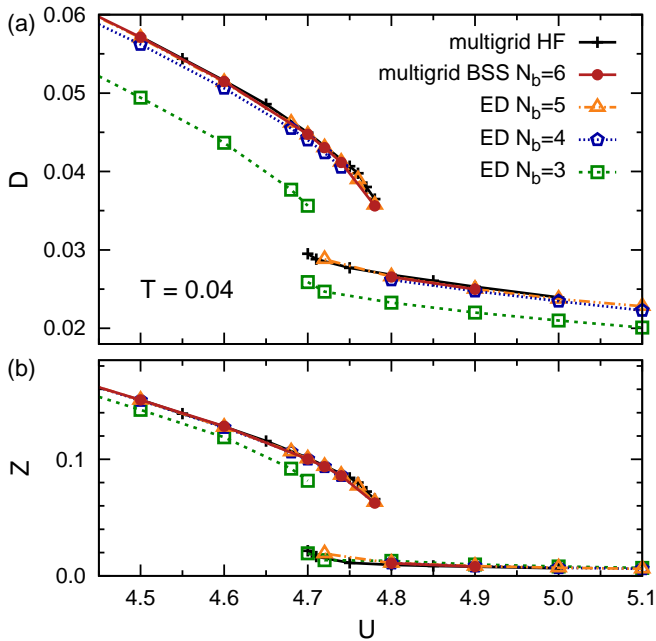}}
  \end{picture}
	\caption{(Color online) Estimates of double occupancy $D(U)$ and quasiparticle weight $Z(U)$ at $T=0.04$, obtained in selfconsistent DMFT calculations using Hamiltonian based impurity solvers with $3\le N_b\le 6$ bath sites: multigrid BSS-QMC (circles), ED (open symbols). Multigrid HF-QMC results (crosses) represent the limit $N_b\to\infty$. In each panel, the upper (lower) sets of curves correspond to metallic (insulating) solutions. Lines are guides to the eye only.}
	\label{fig:D_Z_B25_FS}
\end{figure}
\reffl{D_Z_B25_FS} shows estimates of $D(U)$ and $Z(U)$ at $T=0.04$, similarly to \reff{D_Z_B25} and with the same multigrid HF-QMC reference data (crosses), but now using Hamiltonian based impurity solvers with $3\le N_b\le 6$ bath sites. Here and in the following, ``BSS'' refers to multigrid BSS-QMC data, i.e., without significant Trotter errors; for simplicity, we have used this method only for the largest auxiliary Hamiltonian ($N_b=6$). Smaller bath sizes ($N_b=3$, $N_b=4$, and $N_b=5$) are represented only by the ED solution, which is cheaper and free of statistical noise. At the resolution of \reff{D_Z_B25_FS}, the estimates associated with the finer bath discretizations $N_b=5$ (triangles) and $N_b=6$ (circles) agree with each other. Therefore and since they are also consistent with the unbiased multigrid HF-QMC data (crosses), we conclude that convergence with respect to the bath discretization is reached already at $N_b=5$ at $T=0.04$. In contrast, the ED estimates of $D$ are apparently slightly too small at $N_b=4$ (pentagons); corresponding results at $N_b=3$ (squares) are far off both for $D$ and $Z$. 

Note that consistent convergence of observable estimates with $N_b$, as demonstrated in \reff{D_Z_B25_FS} (as well as \reff{D_Z_B50_FS} and \reff{D_Z_B100_FS}) 
can only be observed when optimal Hamiltonian parameters are determined with great care, as described in \refss{DMFT}, within each self-consistency cycle; otherwise some bath sites may remain ineffective or the estimates can even get worse upon increasing $N_b$. In addition (as always in the DMFT context), it is essential that enough DMFT iterations are performed at each phase point in order to ensure convergency with respect to the self-consistency cycle (cf. \reff{DMFT_cycle}).
\begin{figure} [t] 
  \includegraphics[width=\columnwidth]{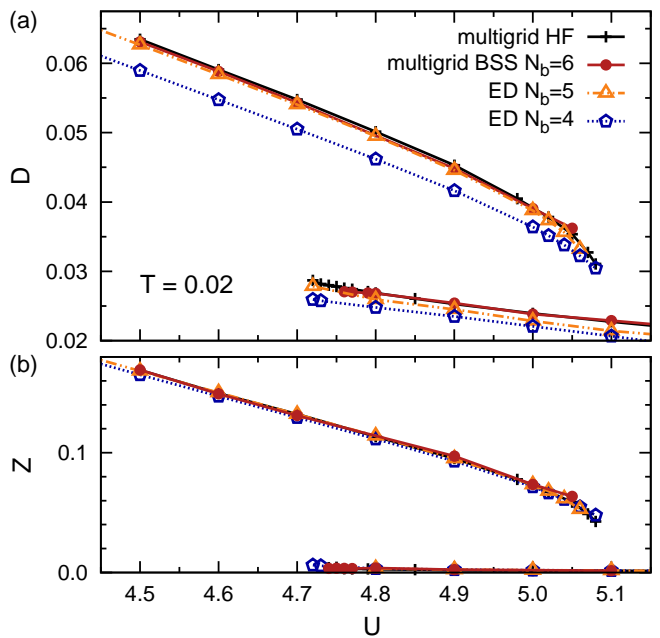}
	\caption{(Color online) Estimates of double occupancy $D(U)$ and quasiparticle weight $Z(U)$ at $T=0.02$, using bath discretizations with $4\le N_b\le 6$ sites, analogous to \reff{D_Z_B25_FS}.}
	\label{fig:D_Z_B50_FS}
\end{figure}

Halving the temperature amplifies the bath discretization effects, as seen in \reff{D_Z_B50_FS}:
At $T=0.02$, only the best Hamiltonian representation ($N_b=6$, evaluated with multigrid BSS-QMC, circles) recovers all reference multigrid HF-QMC results (crosses) within their accuracy. At $N_b=5$, the estimates of $D$ in the insulating phase are already slightly too small; at $N_b=4$, strong negative deviations in $D(U)$ are apparent also for the metallic solution.
\begin{figure} [t] 
  \vspace{2ex}
  \unitlength0.1\columnwidth
  \begin{picture}(10,10)
    \put(0,0){\includegraphics[width=\columnwidth]{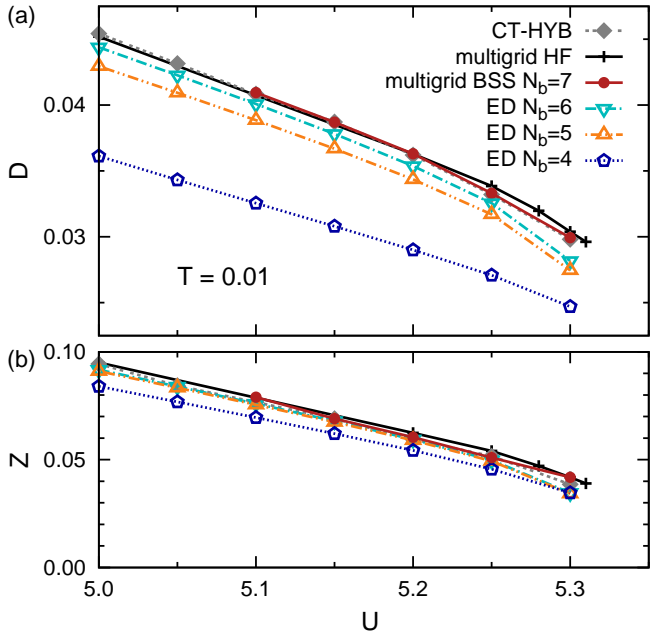}}
  \end{picture}
	\caption{(Color online) Estimates of double occupancy $D(U)$ and quasiparticle weight $Z(U)$ at $T=0.01$, using bath discretizations with $4\le N_b\le 7$ sites, analogous to \reff{D_Z_B25_FS}. The CT-HYB data (diamonds) also represent the limit $N_b\to\infty$.}
	\label{fig:D_Z_B100_FS}
\end{figure}
The impact of the bath discretization becomes even much stronger at $T=0.01$, as shown in \reff{D_Z_B100_FS} for the metallic phase, which is interesting as a strongly renormalized Fermi liquid (while the properties of the insulating phase are asymptotically independent of temperature). We find that even the results for $N_b=5$ and $N_b=6$ deviate significantly in \reffa{D_Z_B100_FS} from each other and from the multigrid HF-QMC reference result, especially at $U=5.3$, near the edge of the stability region of the metallic phase. Only the multigrid BSS-QMC results using $N_b=7$ bath sites (circles) agree with the reference data within their precision; even better agreement is observed with data obtained using the CT-HYB impurity solver (diamonds). Note that BSS-QMC is much more efficient than ED at $N_b=7$; in particular, the latter would need orders of magnitude more main memory than the former.

As noted in \refss{BSS}, a strong increase with inverse temperature of the number $N_b$ of bath sites needed for a given accuracy could, in principle, eliminate the scaling advantage of the DMFT-BSS approach, as its computational cost does not only include the direct factor $\beta$, but also a factor $N_b^3\equiv N_b^3(\beta)$. However, our results indicate that this effect is minor: In our test case, we needed to add one bath site upon halving the temperature for roughly constant accuracy; this is consistent with a scaling $N_b\propto \ln(\beta)$, i.e., an overall computational cost proportional to $\beta \left[\ln(\beta)\right]^3$ which is still linear up to logarithmic corrections. 

\section{Conclusions}\label{sec:conclusion}

The DMFT and its extensions are invaluable tools for the study of phenomena associated with strong electronic correlations and for quantitative predictions of properties of correlated materials. However, the numerical solution of the DMFT self-consistency equations remains a great challenge: the established, direct, QMC impurity solvers yield unbiased results, but
provide only limited access to the low-$T$ phase regions of interest, due to the cubic scaling of their computational cost with the inverse temperature $\beta$. Exact diagonalization (ED) approaches, on the other hand, are limited by their exponential scaling with the number of sites $N$ of the auxiliary Hamiltonian.

The multigrid BSS-QMC algorithm presented in this work allows for solving the DMFT self-consistency equations with an effort that grows only linearly with $\beta$; in contrast to an earlier BSS-QMC based method \cite{Khatami2010a}, it is free of significant Trotter errors, i.e., numerically exact at the level of the auxiliary Hamiltonian. Since the computational cost grows only cubically with $N$, much better representations of the bath are possible than for ED. As demonstrated by applications to the half-filled Hubbard model in and near the coexistence region of metallic and insulating solutions and by comparisons with direct QMC impurity solvers, the new method yields unbiased results (for sufficiently fine bath discretization), 
in spite of using quite coarse Trotter discretizations in the underlying BSS-QMC evaluations.

The new unbiased quasi-CT impurity solver should show its full potential in multi-band cases and in cluster extensions of DMFT, where the prefactor $N^3$ of the BSS-QMC scheme (compared to a factor of $1$ in HF-QMC calculations in single-site DMFT) is leveled off by the increased complexity of the original DMFT problem. Our approach can also be extended beyond Hubbard models; it could be particularly valuable for the cellular DMFT treatment of the Kondo lattice model, where interesting temperature regimes are out of reach of the existing impurity solvers \cite{Werner2006b,KLM1,*KLM2}.

We thank E.\ Gorelik, E.\ Khatami, R.\ Scalettar, and P.\ Werner for valuable discussions. Financial support by the Deutsche Forschungsgemeinschaft through FOR 1346 is gratefully acknowledged.

\bibliography{BSS_DMFT_Multigrid}
\end{document}